\documentclass[final,1p,times,geometry={paperwidth=8.5in,paperheight=11in,margin=1in}]{elsarticle}
\usepackage{amsmath,amssymb}
\usepackage{graphicx}
\usepackage{float}
\usepackage[T1]{fontenc}
\usepackage{ragged2e}
\usepackage{booktabs}
\usepackage[none]{hyphenat}
\journal{Scripta Materialia}
\biboptions{sort&compress}
\newcommand{\ndfeb}{\mathrm{Nd_2Fe_{14}B}}
\newcommand{\invcm}{\ensuremath{\,\mathrm{cm}^{-1}}}

\begin{document}

\begin{frontmatter}

\title{Grain Boundary Distortion Reorients the Local 4\textit{f} Easy Axis and Splits Light from Heavy Lanthanides in $\ndfeb$}

\author[mc]{Avik Mahata\corref{cor}}
\ead{mahataa@merrimack.edu}
\author[nv]{M. Zakotnik}

\cortext[cor]{Corresponding author. Author list provisional.}
\address[mc]{Department of Mechanical and Electrical Engineering, Merrimack College, North Andover, MA, USA}
\address[nv]{Mendelan LLC, Hudson, OH, USA}

\begin{abstract}
The coercivity of rare-earth permanent magnets originates from the crystal field acting on localized $4f$ electrons, yet how grain boundaries modify the underlying single-ion Hamiltonian remains largely unknown. Here we determine the $4f$ crystal field in bulk and grain-boundary environments of $\ndfeb$ using embedded multireference electronic-structure calculations for the substitutional series Ce, Pr, Nd, Gd, Tb, and Dy. Grain-boundary distortion produces a crossover across the lanthanide series: the crystal field weakens by up to 36\% for light lanthanides but strengthens by about 25\% for heavy lanthanides, with Nd marking the crossover. The ground-state doublet of Dy becomes more isolated and axial, whereas Ce exhibits a collapsed low-lying excitation that suppresses anisotropy. The local easy axis rotates from the bulk $c$ axis toward the basal plane, identifying grain-boundary sites as favorable nucleation centers for magnetization reversal. These results provide an atomistic basis for heavy rare-earth grain-boundary diffusion and furnish transferable single-ion parameters for spin-lattice simulations.
\end{abstract}


\end{frontmatter}

Rare earth permanent magnets convert the localized magnetism of the lanthanides into the largest energy products of any material, and they are now indispensable to electrification, powering electric vehicle traction motors, wind turbine generators, and robotic actuators \cite{gutfleisch2011,coey2020,skomski1999}. Among them $\ndfeb$ remains the benchmark: since its discovery it has offered a combination of remanence, coercivity, and cost that no competing chemistry has displaced \cite{sagawa1984,herbst1991}. The field has remained tied to this one compound for structural as much as economic reasons: the tetragonal lattice places the Nd sublattice in a site whose crystal field produces a strong uniaxial anisotropy, and decades of work on rare earth lean and rare earth free chemistries have not matched that anisotropy at comparable magnetization, so replacing $\ndfeb$ has proven far harder than improving it \cite{coey2020,skomski1999}.

Improving it, however, is overwhelmingly a problem of microstructure rather than of the ideal crystal. The measured coercivity of a sintered magnet reaches only a fraction of the ideal anisotropy field, a discrepancy recognized since Brown as the coercivity paradox \cite{brown1945}, and the size of the shortfall is governed by the nucleation and pinning of reverse domains at grain boundaries \cite{kronmuller1987,kronmuller2003}. Every industrial route to high temperature coercivity therefore operates on the boundary region. Heavy 4\textit{f} metals such as Dy and Tb are diffused along the grain boundaries to harden the outer shell of each grain \cite{hono2012,sepehriamin2013,sepehriamin2023}, the neodymium rich intergranular phase is tuned to magnetically decouple neighboring grains \cite{hono2012}, and, increasingly, the feedstock is recovered from end of life magnets, which returns residual Co, Cu, Zr, B, and oxygen that redistribute during hydrogen decrepitation, sintering, and annealing \cite{binnemans2013,yang2017,zakotnik2025a,zakotnik2025b}. Each of these levers changes the local chemical and geometric surroundings of individual 4\textit{f} ions in and near the boundary.

This is where recycling and redesign meet a genuine bind. Recovering coercivity in a recycled $\ndfeb$ magnet typically requires adding back heavy rare earths such as Dy or Pr, so the scarce, strategically constrained ions are the very ones consumed to rescue the recycled product: one spends rare earth to recycle rare earth \cite{binnemans2013,yang2017}. Escaping that loop means designing new or substituted 4\textit{f} alloys whose anisotropy can be anticipated rather than measured after the fact, and here the available modeling frameworks fall short in a specific way. Periodic density functional theory with a Hubbard correction is the standard tool for intermetallics and their defects, but it collapses the 4\textit{f} multiplet onto a single determinant and does not recover the zero temperature anisotropy of $\ndfeb$ unless Hund's rules are imposed by hand, converging to seed dependent solutions when low lying states compete \cite{dudarev1998,lee2025,ke2025,larson2003}. Such calculations largely assume the material is already a magnet and describe how that magnetism responds, rather than resolving where the anisotropy originates. Dynamical mean field theory recovers a bulk crystal field but has been applied to the ideal, single site host \cite{delange2017,amadon2006,eryigit2022}, and a density functional point charge treatment of point defects in a related 1:12 magnet shows that a structurally single phase sample can nonetheless carry a spatially varying anisotropy \cite{patrick2024}. Complementary process level methods do not fill the gap either: CALPHAD modeling of the relevant oxide and alloy systems captures phase equilibria and the thermodynamics of recovery accurately \cite{saunders1998,lukas2007,wei2026a,wei2026b}, but by construction it describes stability and composition and does not return magnetization, anisotropy, or coercivity for a new material. Likewise, atomistic spin lattice dynamics can propagate magnetization on realistic microstructures, yet it requires specially constructed magnetic interatomic potentials that presently describe only an itinerant, collinear moment and omit the rare earth crystal field altogether \cite{tranchida2018,nitol2026,corvacho2026,murillopolo2026,eriksson2017}.

What is missing, then, is a first principles account of the 4\textit{f} single ion Hamiltonian in a real, processed microstructure. Precisely this Hamiltonian is resolved to spectroscopic accuracy by the molecular magnetism community: state averaged complete active space calculations over the seven 4\textit{f} orbitals, corrected for dynamic correlation and spin--orbit coupling and projected onto an effective ligand field, deliver the crystal field coefficients, multiplet energies, and \textit{g} tensors of lanthanide complexes and reproduce their measured susceptibilities \cite{ungur2017,atanasov2012,atanasov2015,aravena2016,angeli2001}. This machinery is mature and built on relativistic basis sets designed for the lanthanides \cite{neese2022,pantazis2009,aravena2016basis}; the same localized, chemically robust 4\textit{f} shell that makes it work underlies rare earth quantum technologies \cite{tittel2025}. Yet it has been turned almost exclusively on isolated molecules, and only rarely on the extended, defect bearing solids that actually constitute a permanent magnet \cite{larson2003,delange2017}. In this Letter we apply that multireference machinery to the rare earth site of $\ndfeb$ in both its bulk and its grain boundary environment, and across a light to heavy substitutional series, to ask directly how the microstructure rewrites the local 4\textit{f} Hamiltonian.

\begin{figure}[H]
  \centering
  \includegraphics[width=\linewidth]{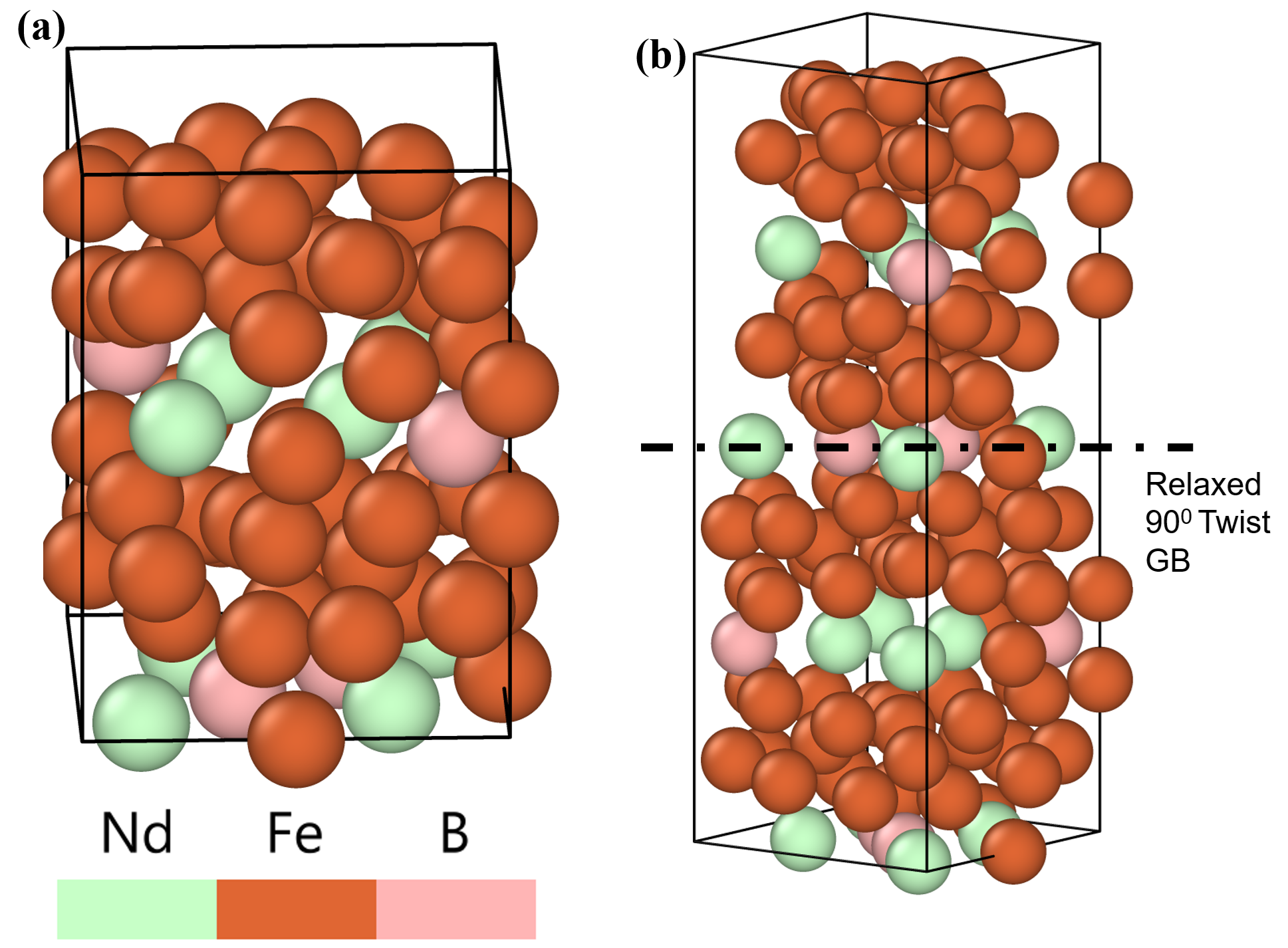}
\caption{Structural models used in this work. (a) Relaxed bulk Nd$_2$Fe$_{14}$B supercell containing 68 atoms. (b) Relaxed 90$^\circ$ twist coincidence-site lattice grain-boundary supercell containing 136 atoms. Rare-earth (Nd), Fe, and B atoms are shown in green, orange, and pink, respectively. The dashed line marks the grain-boundary plane.}
  \label{fig:structure}
\end{figure}

The local $4f$ Hamiltonian at each rare-earth site was resolved using a multireference embedded-cluster approach followed by an \textit{ab initio} ligand-field projection, the same framework that reproduces the electronic structure and magnetic anisotropy of lanthanide single-molecule magnets to spectroscopic accuracy \cite{ungur2017,atanasov2012,aravena2016}. For each trivalent lanthanide (Ce, Pr, Nd, Gd, Tb, and Dy), the electronic structure was built progressively from the free ion to the embedded crystal environment through a three-stage workflow consisting of (i) state-averaged complete active space self-consistent field (SA-CASSCF) calculations on the isolated ion, (ii) embedded SA-CASSCF calculations in the crystal electrostatic field, and (iii) spin--orbit state interaction followed by \textit{ab initio} ligand-field projection to recover the crystal-field Hamiltonian and magnetic observables. The embedding environments were constructed from density functional theory relaxed bulk and grain-boundary structures of $\ndfeb$ obtained in our companion study \cite{mahata2026}. A relaxed 68-atom bulk supercell and a relaxed 136-atom $90^\circ$ twist grain-boundary supercell (Fig.~\ref{fig:structure}) were used to define the local coordination environments. These relaxed structures serve only as structural inputs to the multireference calculations and are not further modified in this work.

For each environment, the target lanthanide ion was treated quantum mechanically together with its first coordination shell, while the remaining crystal was represented by a charge-neutral array of point charges reproducing the Madelung potential. Identical trivalent ions were placed at equivalent bulk and grain-boundary sites so that the calculations differ only in the local coordination geometry and nearest-neighbor environment, thereby isolating the effect of the grain boundary. State-averaged CASSCF calculations were performed over the complete $4f^n$ active space, with the active space fixed by the formal trivalent occupancy, ranging from one electron in seven orbitals for Ce(III) to nine electrons in seven orbitals for Dy(III), and averaging over all spin-free terms of the configuration \cite{roos1980}. Dynamic correlation beyond the active space was recovered using strongly contracted N-electron valence perturbation theory \cite{andersson1992,angeli2001,angeli2002}, followed by spin--orbit coupling through quasidegenerate perturbation theory \cite{malmqvist2002}. The resulting correlated wavefunctions were projected onto an effective one-electron ligand-field Hamiltonian to obtain the seven $4f$ orbital energies, Racah parameters, spin--orbit constant, crystal-field coefficients in the Stevens operator formalism, and the ground-state $g$ tensor \cite{stevens1952,hutchings1964,rudowicz1985,newman2000,jensen1991,atanasov2012,aravena2016,ungur2017,chibotaru2012}. Scalar relativistic effects were included through the second-order Douglas--Kroll--Hess Hamiltonian together with quadruple-$\zeta$ relativistically contracted basis sets developed for the lanthanides \cite{reiher2004,pantazis2009,aravena2016basis}. All multireference calculations were performed with the ORCA electronic structure package \cite{neese2022,neese2012}.

The complete workflow was applied systematically to both bulk and grain-boundary environments for the Ce--Dy substitutional series. The correlated spin--orbit wavefunctions were subsequently used to determine crystal-field splittings, ground-state doublet splittings, isolation gaps, magnetic easy-axis orientations, temperature-dependent susceptibilities ($\chi_MT$), and volumetric $4f$ orbital cube files for visualization. The same correlated electronic structure therefore provides the basis for every quantity reported in this work.

The bulk calculations reproduce the expected crystal-field trends and provide the reference against which the grain-boundary effects are assessed. The total $4f$ ligand-field splitting decreases monotonically from Ce to Dy, from 2589 to 1002\invcm{} (Table~\ref{tab:cf}), reflecting the well-known lanthanide contraction, in which the increasingly contracted $4f$ shell becomes progressively shielded by the $5s5p$ core and experiences a weaker crystal field \cite{aravena2016,freeman1962}. Gadolinium, a half-filled $S$-state ion with $L=0$, follows the same structural trend but carries no orbital anisotropy, providing a useful control that distinguishes changes in the crystal field from changes in the magnetic response. The vanishing Racah parameter $B$ obtained for Ce is consistent with the $f^{1}$ limit of negligible interelectronic repulsion, and the ground multiplet easy axis remains aligned with the crystallographic \textit{c} axis for every bulk environment, consistent with the established uniaxial anisotropy of $\ndfeb$ \cite{hirosawa1986,delange2017}.

\begin{table}[H]
\centering
\caption{Total 4\textit{f} ligand field splitting (spread of the seven orbital energies) at the bulk and boundary sites, the boundary to bulk ratio, the splitting of the lowest orbital pair, and the isolation gap to the next orbital. All energies in $\mathrm{cm}^{-1}$.}
\label{tab:cf}
\small
\begin{tabular}{lccccc}
\toprule
Ion & Bulk & GB & Ratio & Low pair & Isolation \\
    &      &    &       & (bulk$\to$GB) & (bulk$\to$GB) \\
\midrule
Ce & 2589 & 1655 & 0.64 & 406$\to$6 & 1461$\to$768 \\
Pr & 1835 & 1595 & 0.87 & 137$\to$5 & 1177$\to$740 \\
Nd & 1464 & 1513 & 1.03 & 65$\to$4  & 921$\to$703 \\
Gd & 1065 & 1314 & 1.23 & 18$\to$3  & 479$\to$613 \\
Tb & 1041 & 1295 & 1.24 & 16$\to$3  & 465$\to$605 \\
Dy & 1002 & 1261 & 1.26 & 12$\to$3  & 431$\to$589 \\
\bottomrule
\end{tabular}
\end{table}

\begin{figure}[H]
  \centering
  \includegraphics[width=\linewidth]{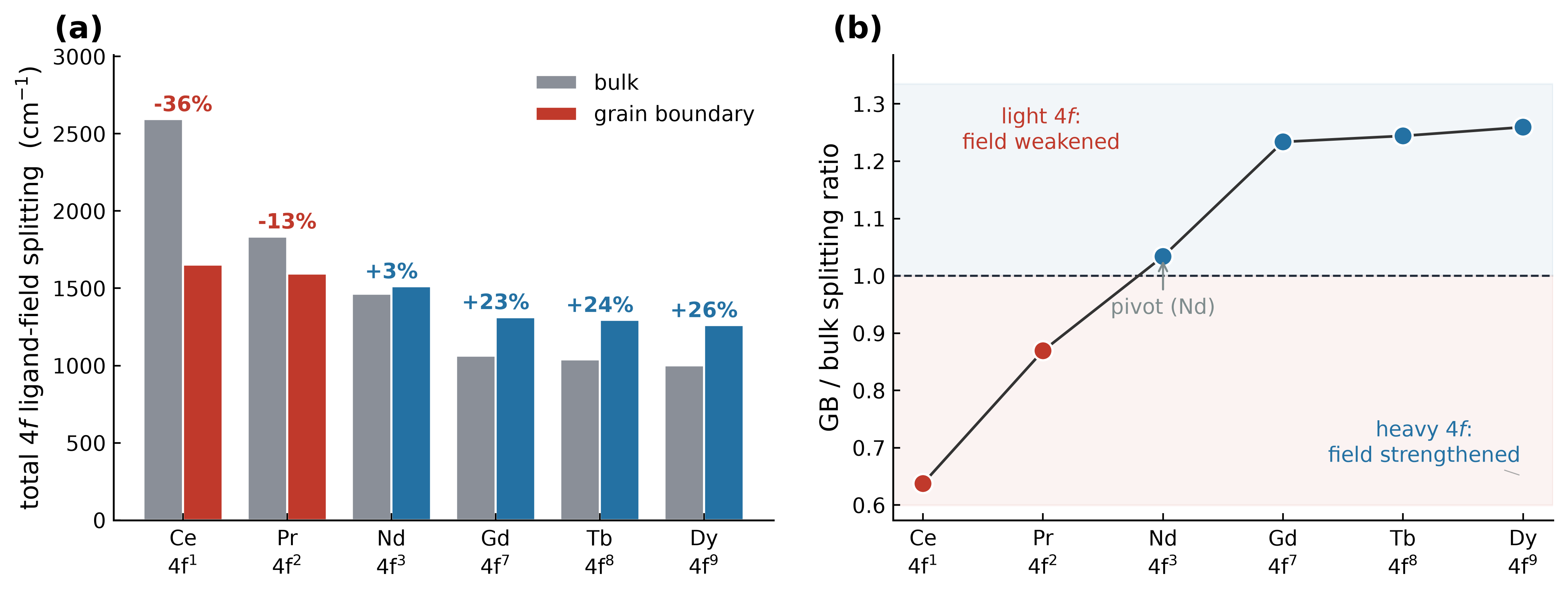}\\[4pt]
  \includegraphics[width=\linewidth]{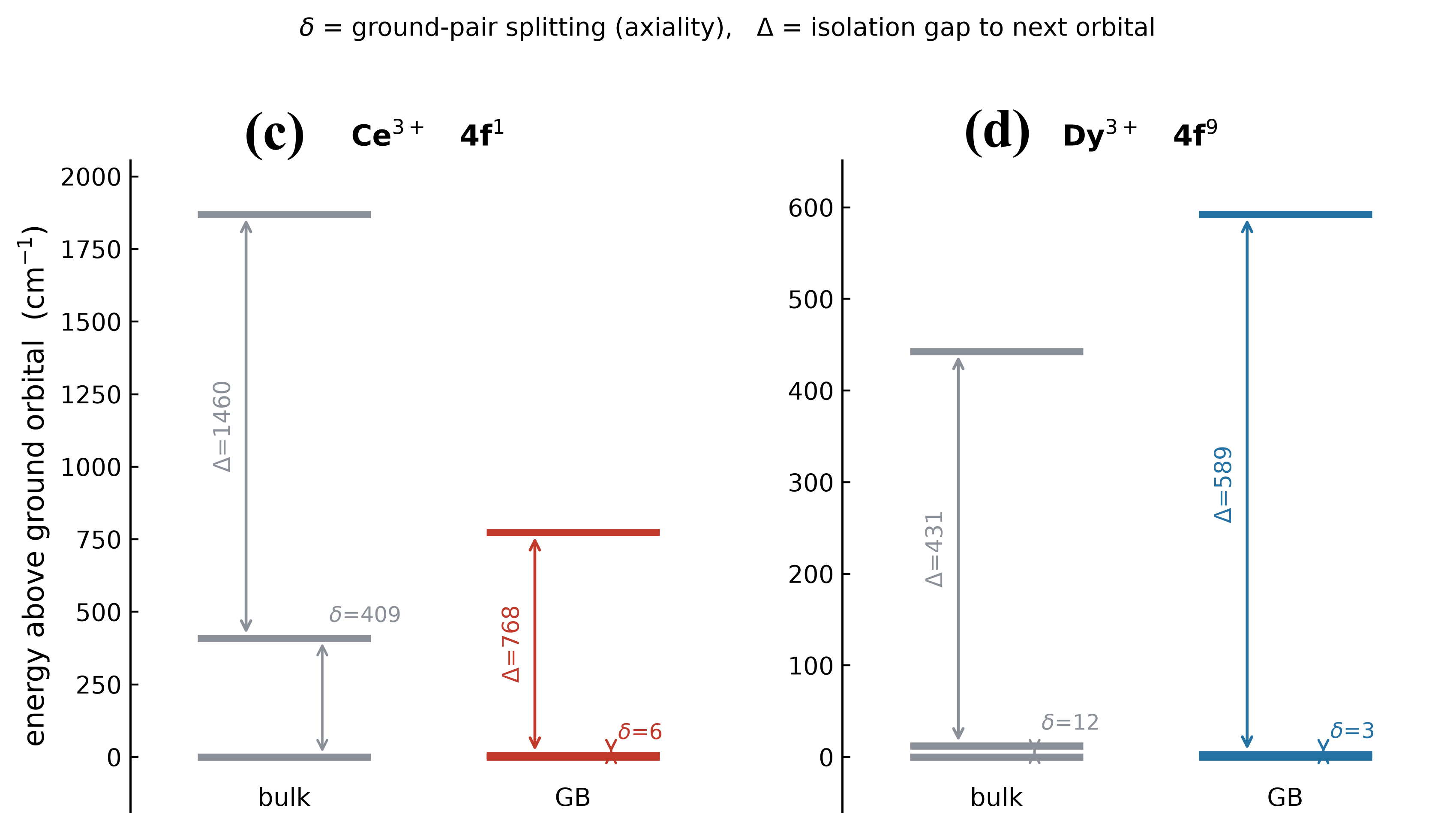}
\caption{Grain-boundary modification of the crystal field. (a) Total $4f$ ligand-field splitting in bulk and grain-boundary (GB) environments. (b) Ratio of GB-to-bulk splitting across the rare-earth series. Crystal-field energy levels for (c) Ce$^{3+}$ ($4f^{1}$) and (d) Dy$^{3+}$ ($4f^{9}$), showing the ground-state splitting ($\delta$) and isolation gap ($\Delta$).}
  \label{fig:crossover}
\end{figure}

The central result of this work is the opposite response of the crystal field to grain-boundary distortion across the lanthanide series. The total ligand-field splitting (Fig.~\ref{fig:crossover}a) decreases substantially for the light lanthanides, falling by 36\% for Ce and 13\% for Pr, whereas it increases by 23--26\% for the heavy lanthanides Gd, Tb, and Dy. The corresponding grain-boundary-to-bulk ratio (Fig.~\ref{fig:crossover}b) crosses unity at Nd, which changes by only 3\% and therefore marks the crossover between the two regimes. The sign of the grain-boundary perturbation is governed primarily by the radial extent of the $4f$ shell rather than by the ground-state charge distribution. The diffuse light-$4f$ orbitals are destabilized by the more open grain-boundary coordination, whereas the more contracted heavy-$4f$ orbitals experience a stronger and more axially organized crystal field.

The microscopic origin of this crossover is illustrated by the representative crystal-field level schemes for Ce and Dy (Fig.~\ref{fig:crossover}c,d). The representative crystal-field level schemes for Ce and Dy (Fig.~\ref{fig:crossover}c,d) illustrate how the grain boundary drives the lowest orbital pair toward near degeneracy while modifying the isolation gap in opposite ways for light and heavy lanthanides. In Dy (Fig.~\ref{fig:crossover}d), the ground-pair splitting decreases from 12 to 3\invcm{}, while the isolation gap to the next orbital increases from 431 to 589\invcm{}. A nearly degenerate and well-isolated ground doublet is characteristic of a robust axial anisotropy, demonstrating that the grain boundary reinforces the magnetic character of the heavy ion precisely where Dy is introduced through grain-boundary diffusion \cite{hono2012,sepehriamin2013,patrick2024}. Ce exhibits the opposite behavior (Fig.~\ref{fig:crossover}c). The first excited orbital approaches the ground pair, reducing the isolation gap from 1461 to 768\invcm{}, thereby enhancing state mixing and suppressing an already weak anisotropy. Together, these results provide a microscopic single-ion explanation for why heavy rare-earth grain-boundary diffusion enhances coercivity, whereas light rare-earth substitution does not.

\begin{figure}[H]
  \centering
  \includegraphics[width=\linewidth]{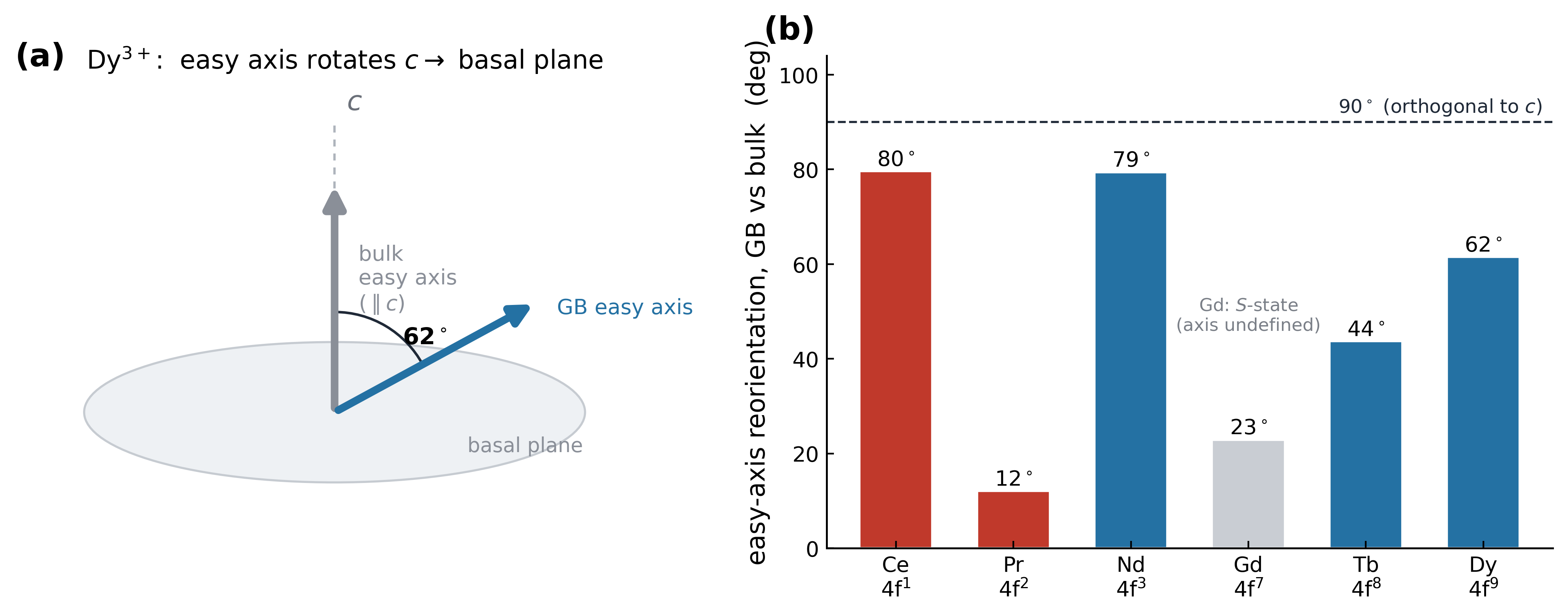}
\caption{Grain-boundary-induced magnetic anisotropy. (a) Schematic illustrating the rotation of the magnetic easy axis at the grain boundary. (b) Angle between the bulk and grain-boundary easy-axis directions for the rare-earth ions considered.}
  \label{fig:chit}
\end{figure}

The modified crystal field also changes the orientation of the local magnetic anisotropy. The schematic in Fig.~\ref{fig:chit}a illustrates the rotation of the magnetic easy axis from the bulk \textit{c} direction toward the basal plane at the grain boundary. The calculated reorientation angles for the complete lanthanide series are summarized in Fig.~\ref{fig:chit}b. Although the magnitude of the rotation varies among the ions, every anisotropic ion exhibits a substantial deviation from the bulk easy axis, with Ce and Nd approaching an almost orthogonal orientation. Gadolinium again serves as the control because its orbital angular momentum vanishes, leaving the easy-axis direction ill defined. This reorientation provides a direct connection between the local crystal field and the micromagnetic behavior of the material. A grain-boundary ion whose easy axis is nearly orthogonal to the magnetization of the grain interior is naturally oriented to nucleate a reverse magnetic domain, which is the microscopic origin of nucleation-controlled coercivity \cite{brown1945,kronmuller1987,kronmuller2003}. The geometric rotation is common across the series, whereas its magnetic consequences depend on the crystal-field crossover described above.

\begin{figure}[H]
  \centering
  \includegraphics[width=\linewidth]{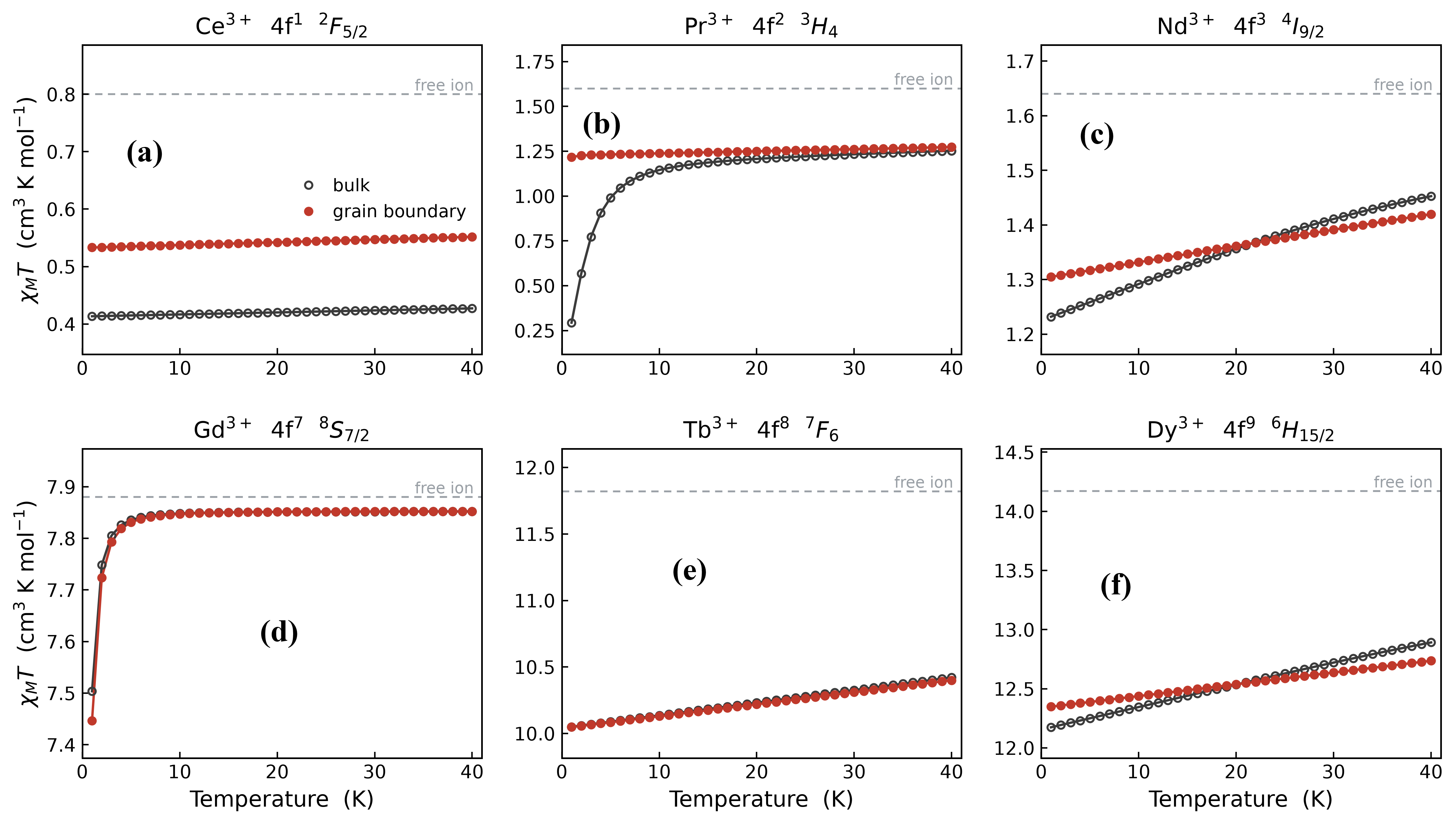}
\caption{Calculated temperature dependence of $\chi_MT$ for (a) Ce$^{3+}$, (b) Pr$^{3+}$, (c) Nd$^{3+}$, (d) Gd$^{3+}$, (e) Tb$^{3+}$, and (f) Dy$^{3+}$ in bulk and grain-boundary environments. Dashed lines denote the corresponding free-ion values.}
  \label{fig:easyaxis}
\end{figure}

The correlated magnetic susceptibility provides an experimentally accessible signature of these crystal-field modifications (Fig.~\ref{fig:easyaxis}). The largest differences between bulk and grain-boundary environments occur at low temperature, where only the lowest crystal-field levels contribute to the magnetic response. Pr provides the clearest example (Fig.~\ref{fig:easyaxis}b): the bulk susceptibility decreases rapidly toward zero, consistent with a nonmagnetic singlet ground state, whereas the grain-boundary calculation retains a finite susceptibility, indicating a paramagnetic ground state. Dy (Fig.~\ref{fig:easyaxis}f) exhibits an enhanced low-temperature susceptibility consistent with its more isolated ground doublet, while Tb changes only modestly. Gadolinium (Fig.~\ref{fig:easyaxis}d) remains essentially unchanged, confirming that the observed differences arise from crystal-field anisotropy rather than structural distortion alone. Because $\chi_MT$ is directly accessible by SQUID magnetometry, these calculations provide a clear route for experimental validation.

\begin{figure}[H]
  \centering
  \includegraphics[width=\linewidth]{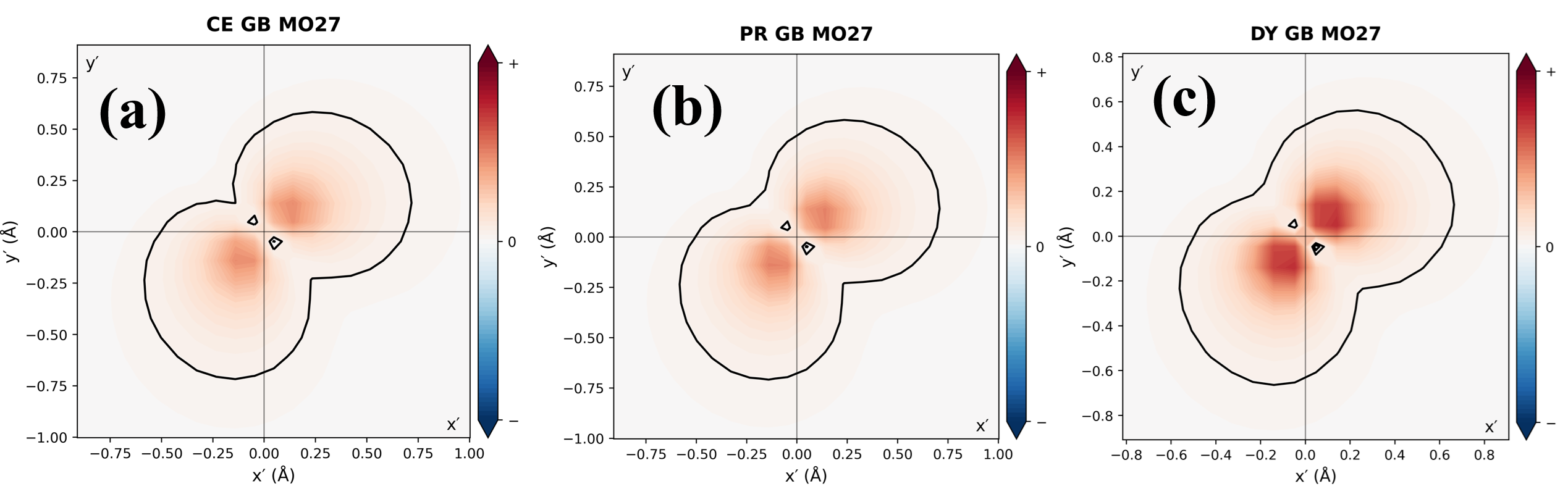}
\caption{Crystal-field potential projected onto the local $x'$--$y'$ plane for (a) Ce, (b) Pr, and (c) Dy at the grain-boundary site. Black contours indicate the local coordination environment, and the color scale represents the crystal-field potential.}
  \label{fig:orbital}
\end{figure}

The real-space origin of these trends is evident in the correlated $4f$ orbitals (Fig.~\ref{fig:orbital}). The bulk ground-state orbital is elongated along the crystallographic \textit{c} axis, consistent with the uniaxial crystal field of the host. At the grain boundary, the orbital reorients toward the basal plane in response to the modified ligand environment. For Ce (Fig.~\ref{fig:orbital}a) and Pr (Fig.~\ref{fig:orbital}b), the orbital density expands into the more open coordination environment, consistent with the reduced crystal-field strength. In contrast, Dy (Fig.~\ref{fig:orbital}c) exhibits a more localized orbital confined by the stronger equatorial crystal field. These orbital densities therefore provide the real-space origin of the crystal-field crossover and demonstrate that the easy-axis rotation reflects a redistribution of the $4f$ charge rather than a change in the ion itself. These environment-resolved crystal-field Hamiltonians provide the missing single-ion input required for realistic atomistic spin-lattice simulations of rare-earth permanent magnets, where existing magnetic interatomic potentials describe only itinerant magnetic moments and neglect the localized $4f$ crystal field \cite{tranchida2018,nitol2026,corvacho2026,murillopolo2026,eriksson2017}. More broadly, this work establishes a multireference framework for resolving environment-dependent crystal fields in realistic permanent-magnet microstructures, providing a predictive route toward crystal-field engineering, grain-boundary design, and multiscale spin-lattice simulations.

Multireference calculations of the local $4f$ Hamiltonian in bulk and grain-boundary environments of $\ndfeb$ show that the electronic structure of rare-earth ions is governed by the local microstructure rather than by a fixed atomic property. Grain-boundary distortion produces a crossover across the lanthanide series, weakening the crystal field for the light lanthanides while strengthening and axially sharpening it for the contracted heavy lanthanides. At the same time, the local magnetic easy axis rotates away from the bulk \textit{c} direction toward the basal plane, identifying grain-boundary sites as favorable locations for magnetization reversal. These trends provide a microscopic single-ion explanation for the effectiveness of heavy rare-earth grain-boundary diffusion and offer experimentally testable predictions through the calculated magnetic susceptibilities. More broadly, the environment-resolved Hamiltonians obtained here provide transferable single-ion parameters for spin-lattice simulations and establish a multireference framework for predicting the influence of realistic microstructural environments on crystal fields and magnetic anisotropy in rare-earth permanent magnets.

\section*{Data availability}
The relaxed structures and the correlated output files are available from the authors on reasonable request.

\section*{Acknowledgements}
This work was supported by the Department of Mechanical and Electrical Engineering at Merrimack College. The author acknowledges the use of computational resources at the Massachusetts Green High Performance Computing Center (MGHPCC). This research also benefited from high performance computing allocations provided by the National Science Foundation through ACCESS (awards MAT250103 and MAT240094). Additional computational resources were supported by Argonne National Laboratory under the Director's Discretionary allocation for the project \textit{AIAlloyLW}. Further support was provided through a National Science Foundation MRI Award to Wilkes University (Award No.\ 1920129), which contributed essential computational infrastructure for this study.

\bibliographystyle{elsarticle-num}
\bibliography{references}

\end{document}